\documentclass[aps,prb,twocolumn,
               showpacs,amsmath,amssymb,superscriptaddress]{revtex4}

\usepackage{amsfonts}
\usepackage{amsmath}
\usepackage{dcolumn}
\usepackage[dvips]{graphicx,color}
\begin{document}

\title{The Crossover from Impurity to Valence Band 
       in Diluted Magnetic Semiconductors: \\
The Role of the Coulomb Attraction by Acceptors}

\author{F. Popescu}
\email{pf02@garnet.acns.fsu.edu} 
\affiliation{Department of Physics,
             Florida State University, Tallahassee, FL 32306}

\author{C. \c{S}en}
\affiliation{Department of Physics, 
             Florida State University,Tallahassee, FL 32306} 
\affiliation{National High Magnetic Field Laboratory, Tallahassee, FL 32310}

\author{E. Dagotto}
\affiliation{Department of Physics and Astronomy, University of
Tennessee, Knoxville, TN 37996}\affiliation{Materials Science and
Technology Division, Oak Ridge National Laboratory,  Oak Ridge, TN
32831}

\author{A. Moreo}
\affiliation{Department of Physics and Astronomy, University of
Tennessee, Knoxville, TN 37996}\affiliation{Materials Science and
Technology Division, Oak Ridge National Laboratory,  Oak Ridge, TN
32831}

\begin{abstract}
The crossover between an impurity band (IB) and a valence band (VB)
regime as a function of the magnetic impurity concentration in
models for diluted magnetic semiconductors (DMS) is studied
systematically by taking into consideration the Coulomb attraction
 between the carriers and the magnetic impurities. The density of states
and the ferromagnetic transition temperature of a Spin-Fermion model
applied to DMS are evaluated using Dynamical Mean-Field Theory
(DMFT) and Monte Carlo (MC) calculations. It is shown that the
addition of a square-well-like attractive potential can generate an
IB at small enough Mn doping $x$ for values of the $p$-$d$ exchange
$J$ that are not strong enough to generate one by themselves. We
observe that the IB merges with the VB when $x \geqslant x_c$
where $x_c$ is a function of $J$ and the Coulomb attraction strength
$V$. Using MC calculations, we demonstrate that the range of the
Coulomb attraction plays an important role. While the on-site
attraction, that has been used in previous numerical simulations,
effectively renormalizes $J$ for all values of $x$, an unphysical
result, a nearest-neighbor range attraction renormalizes $J$ only at
very low dopings, i.e., until the bound holes wave functions start
to overlap. Thus, our results indicate that the Coulomb attraction
can be neglected to study Mn doped GaSb, GaAs, and GaP in the
relevant doping regimes, but it should be included in the case of Mn
doped GaN that is expected to be in the IB regime.
\end{abstract}

\pacs{71.10.-w, 75.50.Pp.}

\maketitle

\section{Introduction}\label{Section1}

The development of spintronics devices \cite{zutic} has motivated
a large body of research on diluted magnetic semiconductors
\cite{OHN96,dietl} with the ultimate aim of creating materials
with Curie temperatures ($T_{\mathrm{C}}$) above room temperature.
This ambitious goal can only be achieved by a detailed understanding of the
underlying mechanisms that govern the behavior of currently available DMS.

Most theoretical approaches to study these materials start with one
of two extreme regimes: \textit{(i)} the limit of high Mn doping in
which holes are directly doped into the valence band and, thus, are
uniformly distributed in the sample (VB scenario)
\cite{OHN96,dietl,macdonald} and \textit{(ii)} the limit of very low
Mn doping in which holes are electrically bound to the impurity
cores and an impurity band develops due to wave function overlap as
the number of holes increases (IB scenario).\cite{IB} Researchers
using the VB limit claim that it is valid for all the relevant
dopings, namely $x$$>$$1\%$ in Ga$_{1-x}$Mn$_x$As, and some
experimental results support their view.\cite{Potash,KU03} However,
a similar claim is advanced by the groups promoting the IB scenario,
i.e. that the IB exists up to the largest value of $x$ that has been
reached experimentally ($x\approx 10\%$). This view also appears
supported by the analysis of some experimental
data.\cite{BUR06,oka1}

To solve this apparent puzzle, it is very important to study
theoretically the DMS problem using unbiased techniques that provide
reliable estimations for the value of $x$ where the IB to VB
crossover takes place. Such unbiased approaches could be provided by
numerical techniques: in fact, the MC and DMFT methods have already
been applied to a variety of phenomenological models for the
DMS.\cite{nosotros,POP06,FIS03,TAK03,HWANG05} These previous studies
have been able to determine a crossover between the VB and IB
behaviors, but only as a function of increasing values of the
$p$-$d$ exchange $J$. However, most experimental results appear to
indicate that the realistic  $J$ for (Ga,Mn)As is approximately
1~eV,\cite{OKO98} which corresponds to the weak coupling regime in
which no IB is generated by $J$ alone. In fact, recent results
obtained applying MC techniques to a six-orbital microscopic model,
in which both the correct lattice geometry and the spin-orbit
interactions were considered, indicate that (Ga,Mn)As is indeed in
the VB regime for $x\gtrsim 3\%$.\cite{YIL07} In addition, DMFT
techniques, which allow for the study of the very diluted ($x\ll 1$)
regime, have shown that for values of $J$ in the weak coupling
region, an IB never develops as a function of
$x$.\cite{FIS03,TAK03,HWANG05,POP06} However, experiments based on
electron paramagnetic resonance,\cite{EPR} infrared
spectroscopy,\cite{IR} and magnetization measurements\cite{MM} of
the electronic structure of one Mn ion doped in GaAs have actually
shown the existence of a shallow hole state with binding energy
$\mathrm{E_{b}}$$=$$112.4$ meV centered at the $S$=$5/2$ Mn ion.
Moreover, analytical studies indicated that $\mathrm{E_{b}}$ has
contributions from $both$ the spin-dependent $p$-$d$ hybridization
and the Coulomb attraction between the hole and the ${\mathrm{Mn}}$
trapping center.\cite{BAT00} When additional Mn ions are added, the
wave functions of the bounded holes will start to overlap and an IB
will develop. Further increasing $x$ should widen the IB, locating
it closer to the VB and eventually a regime of complete
hybridization with the holes doped into the VB is expected to occur.
Thus, it is clear that a crossover from the IB to the VB regime
should take place in (Ga,Mn)As as a function of $x$.

In this paper, it will be argued that an IB-VB crossover will be
missed in theoretical studies of materials with a weak $J$ if the
Coulomb attraction is disregarded, while materials with very strong
$J$ will be in the IB regime regardless of doping. In fact, here we
explicitly show that by the simultaneous consideration of $J$ and
$V$ in the formalism, the experimentally observed transition from IB
to VB with increasing $x$ can be understood. The organization of the
paper is the following: in Section \ref{Section2} the non-magnetic
interactions in DMS are described; the model used and the DMFT
technique are presented in Section \ref{Section3}; in Section
\ref{Section4} the results, including MC simulations, are discussed,
and Section \ref{Section5} is devoted to the conclusions.

\section{Spin-Independent Interactions between holes and Magnetic Impurities}
\label{Section2}

As remarked in the Introduction, most of the numerical work on DMS
has been performed on models that focused on the role of the spin
dependent $p$-$d$ exchange $J$ interaction between the spins of the
localized impurities and the doped holes.\cite{nosotros,POP06,FIS03}
This is certainly sufficient to capture qualitatively many of the
properties of these compounds, including the generation of
ferromagnetism. However, non-magnetic interactions between holes and
impurities must  be considered in order to improve the quantitative
agreement with experiments. This additional potential term in the
model has been generally referred to as ``chemical
disorder''($V$),\cite{TWOR94} and it summarizes all the non-magnetic
interactions between the localized impurities and the holes. In this
context, Tworzydlo \cite{TWOR94} used a short range potential (less
than nearest-neighbors range) with a square-well form of depth
$V_0$, and considered both positive (repulsive) and negative
(attractive) values of $V_0$. The potential was introduced to
explain an apparent $x$-dependence of the $p$-$d$ exchange in
Cd$_{1-x}$Mn$_x$S. Dietl \cite{DIETL07} recently used the same
approach to address apparently contradictory experimental results
for Ga$_{1-x}$Mn$_x$N. He also pointed out \cite{DIETL02} that this
kind of extra potential term leds to a chemical shift in the
standard impurity language, or to a valence-band offset in the alloy
nomenclature, and that $J$ and $V$ are actually related
\cite{DIETL07,DIETL92} through the expression $V/J=5(U_{\rm
eff}+2\epsilon_d)/4U_{\rm eff}$ where $U_{\rm eff}$ is an effective
correlation energy for the 3$d$ shell, and $\epsilon_d$ is its
energetic position with respect to the top of the valence band.
However, the value of $V$ is not easy to determine and, thus, it has
been added as an extra free parameter by some authors (with $V$
allowed to take both positive and negative
values).\cite{MIL02,HWANG05,Calde} Other efforts focused just on the
attractive Coulomb interaction between the holes and the
impurities.\cite{BAT00,YANG03,TAK03}

Only some of the previously mentioned investigations have attempted
to study the effects of the Coulomb attraction at finite $x$ with
unbiased techniques. The authors of Ref.\ [\onlinecite{BAT00}]
studied the case of a single Mn impurity, considering the long-range
Coulomb potential supplemented by a central cell correction with a
gaussian or square-well shape, that is routinely introduced in
calculations of bound state energies for impurities in
semiconductors.\cite{pante} For higher dopings, it is believed that
the most important coulombic term is the central-cell contribution
since the long-range potential is screened. In Ref.\
[\onlinecite{TAK03}], the coherent potential approximation (CPA),
very similar in spirit to DMFT, was applied to a single orbital
model which included both the spin dependent $p$-$d$ hybridization
$J$ and an on-site central-cell Coulomb attraction $V$. It was
claimed that the IB-VB crossover for (Ga,Mn)As using $V$=0.6~eV
(chosen to reproduce, in combination with $J$=0.8~eV, the single
impurity bound state energy) should occur for $x \sim 1-3\%$. In
Ref.\ [\onlinecite{Calde}], a repulsive on-site potential was added.
Both the repulsive and attractive cases were considered in Ref.\
[\onlinecite{HWANG05}]. However, these important previous efforts
did not present a systematic analysis of results as a function of
$J$, $V$, and $x$, which is part of the goals of the present study.

In this work we apply DMFT to a model that includes $J$ and the
Coulomb attraction $V$. The density of states (DOS) and $T_{\rm C}$
are studied in a wide range of couplings, hoppings, carrier fillings
$p$, and Mn concentrations $x$, and estimations of the most
appropriate values for different materials are made.  We obtain the
IB-VB crossover for a large class of DMS's and show that with a
suitable strength $V$ included, the IB regime can always be reached
by decreasing the Mn concentration.

\section {Model and DMFT Formalism}\label{Section3}

The Spin-Fermion Hamiltonian used here and in several previous
studies contains a kinetic $t$-term that describes the hopping of
holes between two neighboring $i$ and $j$ lattice sites ($t$ is set
to $1$ to define the energy unit), an exchange interaction (EI)
$J_{H}$-term that anti-aligns the carrier's spin with the magnetic
moment of the impurity (considered classical) at site $I$, and a
$V$-term that takes into account the on-site central-cell part of
the attractive Coulomb potential,\cite{well}
\begin{equation}\label{ham}
{\mathcal{H}}\!=-t\sum_{\langle ij\rangle,\alpha}\!\!
(c^{\dag}_{i\alpha}c_{j\alpha}\!+\textrm{H.c.})+2J_{H}\sum_{I}
\mathbf{S}_{I}\cdot\mathbf{s}_{I}-V\sum_{I}n_{I}.
\end{equation}
Here, $c^{\dag}_{i\alpha}$ ($c_{i\alpha}$) is the creation
(destruction) operator for a hole with spin $\alpha$ at site $i$,
$\mathbf{s}_{i}$=$c^{\dag}_{i\alpha}\mathbf{\sigma}_{\alpha\beta}c_{i\beta}/2$
is the hole's spin, $\mathbf{S}_{I}$=$S\mathbf{m}_{I}$ is the
classical spin of the local moment, and $n_{I}$ is the number of
holes at $I$.

Several details on the DMFT calculations were already presented in
Ref.\ [\onlinecite{POP06}] for the case V=0, thus here only a brief
summary is given and the modifications introduced by a non-zero V
are remarked. DMFT uses the momentum independence of the self-energy
in infinite dimensions
[$\Sigma(\mathbf{p},i\omega_{n})$$\rightarrow$$\Sigma(i\omega_{n})$,
$\omega_{n}$=$(2n\!+\!1)\pi T$] \cite{MUL89} and reproduces the
physics of diluted correlated systems in lower
dimensions.\cite{GEO89} Within DMFT, the bare Green's function
${{G}}_{0}(i\omega_{n})$ contains all the information about the
hopping of carriers onto and off magnetic (with probability $x$) and
nonmagnetic (with probability $1$-$x$) sites. With (\ref{ham}) the
full Green's function ${{G}}(i\omega_{n})$ is solved by integration
obtaining the result:
$\langle{{G}}(i\omega_{n})\rangle$=$x\langle[{{G}}^{-1}_{0}
(i\omega_{n})\!+J{\mathbf{m}}\hat{\sigma}+V\hat{\mathbf{I}}]^{-1}\rangle$+$(1\smash{-}x)\langle{{G}}_{0}(i\omega_{n})\rangle$,
where $J$=$J_{H}S$. \cite{DMFT} This equation, complemented with the
relation $\langle
G^{-1}_{0}(i\omega_{n})\rangle$=$z_{n}\smash{-}(W^{2}/16)\langle
G(i\omega_{n})\rangle$ valid within the assumption of a Bethe
lattice,\cite{FUR94} can be solved with a semicircular
noninteracting $\mathrm{DOS}(\omega)$=$2{\rm
Re}\sqrt{\smash[b]{(W/2)^{2}\smash{-}\omega^{2}}}/\pi W$
($z_{n}$=$\mu$+$i\omega_{n}$, $\mu$ is the chemical potential, and
$W$=$4t$ is the bandwidth). Being spin diagonal, $\langle
G_{0}\rangle$ and $\langle G^{-1}_{0}\rangle$ are expanded in powers
of $\sigma_{z}$ as:
$\langle{\alpha}\rangle$=$\alpha_{0}\hat{\mathbf{I}}$+$\sum_{k}\alpha_{k}\sigma^{k}_{z}$,
where $\alpha_{k}$$\sim$$M^{k}$, $M$ being the order parameter used
to detect the FM transition. To linear order in $M$ we write
$\langle{{G}}^{-1}_{0}(i\omega_{n})\rangle$=$B(i\omega_{n})\hat{\mathbf{I}}$
+$Q(i\omega_{n})\sigma_{z}$ and then $B(i\omega_{n})$ is found from
a $4$-th order equation,
\begin{equation}
B_{\pm}=z_{n}-x\frac{W^{2}}{16}\frac{[B_{\pm}+V\pm
JM]}{[B_{\pm}+V]^{2}-J^{2}}
\!-\!(1\smash{-}x)\frac{W^{2}}{16}\frac{1}{B_{\pm}},\label{RE}
\end{equation}
that at $\mu$=$0$ and with $i\omega_{n}$$\rightarrow$$\omega$ gives
us the low-temperature interacting
$\mathrm{DOS}_{\pm}(\omega)$=$-\mathrm{Im}[B_{\pm}(\omega)]/\pi$ for
up ($+$) and down ($-$) spin configurations.\cite{DOS} The
expression for $Q(i\omega_{n})$:
\begin{eqnarray}
Q&=&x\frac{W^{2}}{16}\left\{\frac{Q\smash{+}J
M}{(B\smash{+}V)^{2}\smash{-}J^{2}}
+\frac{2J^{2}Q/3}{[(B\smash{+}V)^{2}\smash{-}J^{2}]^{2}}\right\}\nonumber\\
&{}&+(1-x)\frac{W^{2}}{16}\frac{Q}{B^{2}},\label{QU}
\end{eqnarray}
leads us to an implicit equation for $T_{\rm C}$ in the form:
\begin{widetext}
\begin{equation}
-\sum_{n=0}^{\infty}\frac{4xW^{2}J^{2}B^{2}}
{[48B^{2}-3(1\smash{-}x)W^{2}]\{[B\smash{+}V]^{2}\smash{-}J^{2}\}^{2}
-3xW^{2}B^{2}\{[B\smash{+}V]^{2}\smash{-}J^{2}\}
-2xW^{2}J^{2}B^{2}}=1,\label{TC}
\end{equation}
\end{widetext}
where $B(i\omega_{n})$ is given by Eq.\ (\ref{RE}) at $M$=$0$. 
The $T_{\rm C}$ contained in $\omega_{n}$ can be obtained from Eq.\
(\ref{TC}) numerically.

\section{Results}\label{Section4}
\subsection{General Analysis}\label{general}

\begin{figure}[t]
{\scalebox{0.33}{\includegraphics[clip,angle=0]{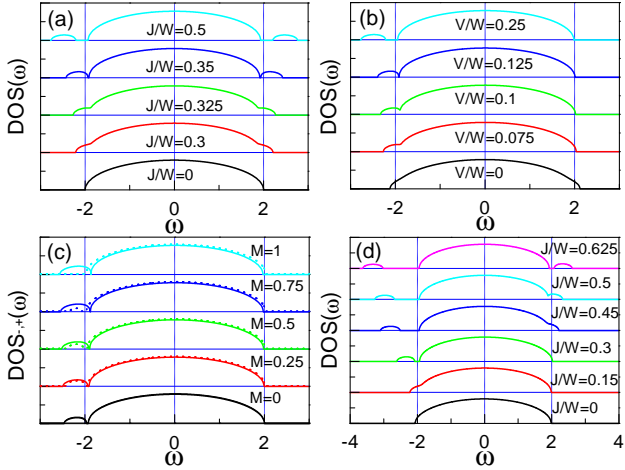}}}
\caption{(Color online) (a) DMFT low-temperature $\mathrm{DOS}$ at
$V$=$0$, $M$=$0$, and different values of $J/W$. An IB forms if
$J/W$ exceeds a critical value $\approx$$0.35$. (b) $\mathrm{DOS}$
at $M$=$0$, $J/W$=$0.25$ (believed to be realistic for (Ga,Mn)As),
and different values of $V/W$. An IB forms if
$V/W$$\geqslant$$0.125$. (c) same as in (b) but at $V/W$=$0.125$ and
for several values of $M$. The solid curve corresponds to
$\mathrm{DOS}_{-}$ while the dotted curve is for $\mathrm{DOS}_{+}$.
(d) $\mathrm{DOS}$ at $M$=$0$, $V/W$=$0.15$, and various $J/W$. With
a $V/W$$\neq$$0$ the electron-hole symmetry is lost. In all frames
the $\mathrm{DOS}$ is in arbitrary units and $x$=$0.035$. At
$x$=$0.05$ we have reproduced the DOS obtained in Ref.\
[\onlinecite{TAK03}] with CPA.} \label{fig1}
\end{figure}

Let us start the discussion of results by considering the general
dependence of a variety of quantities with the parameters of the
model. The DOS obtained from Eq.~(\ref{RE}) at $x$=$0.035$ is
displayed in Fig.~\ref{fig1} for various values of $J$, $M$, and
$V$. As observed in Fig.~\ref{fig1}(a), the $J$-term alone  is able
to generate an IB but only if $J/W$ exceeds a critical value $J_{\rm
c}/W$$\sim$$0.35$. At realistic couplings for (Ga,Mn)As (namely,
$J/W$$\cong$$0.25$ if we assume $J$$\approx$$t$$\sim$$1$eV) there is
no IB generated by the $J$-term alone. However, with the addition of
Coulomb attraction, when a value $V/W$$\geqslant$$0.125$ is reached,
then a well-defined split IB forms, as shown in Fig.\ref{fig1}(b).
No ``symmetric'' impurity band exists at high energies since the
observed one is due to the carriers that are trapped in the vicinity
of the core spins through the influence of $V$, and are fully
aligned for $M$=$1$ (Fig.~\ref{fig1}(c)). The growth of $J/W$
produces asymmetric low- and high- energy impurity bands if
$V$$\neq$$0$ (Fig.~\ref{fig1}(d)).
\begin{figure}[t]
{\scalebox{0.33}{\includegraphics[clip,angle=0]{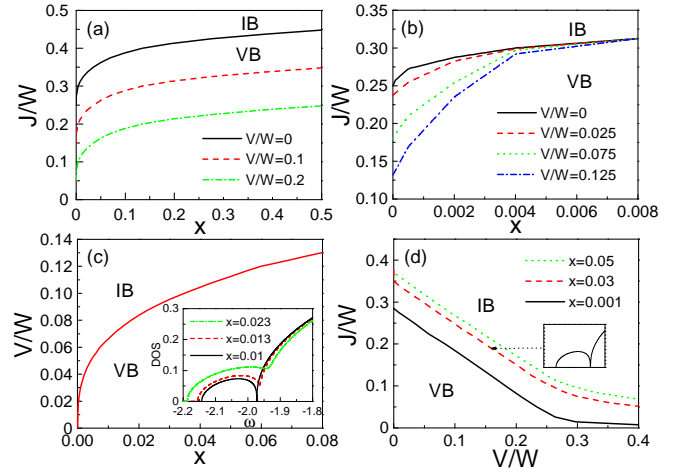}}}
\caption{(Color online) (a) The diagram $J/W$ vs. $x$ for various
values of $V$. The solid curve defines the IB-VB crossover at $V=0$.
(b) The diagram $J/W$ vs. $x$ for (Ga,Mn)As when $V$ is $x$
dependent. The $V$$\neq$$0$ curves all join at $x$$\approx$$0.005$,
that marks the Mn doping concentration beyond which the Coulomb
attraction is no longer relevant and the IB disappears for realistic
couplings. (c) The diagram $V/W$ vs. $x$ at a realistic ratio for
(Ga,Mn)As $J/W$=$0.25$ with an on-site Coulomb attraction. The inset
shows the merging of the impurity and valence bands with increasing
$x$, at $V/W$=$0.066$. (d) The diagram $J/W$ vs. $V/W$ at various
$x$. 
The inset shows the DOS at $J/W$=$0.2$, $V/W$=$0.148$, and
$x$=$0.03$. Since $J_{c}/W$ is $x$-dependent, the VB ``triangle''
shrinks (expands) as $x$ decreases (increases), with the shrinking
saturating at $J_{c}/W\rightarrow$$0.25$.}\label{fig2}
\end{figure}

We have observed that the coupling strength $J_{c}/W$ for which the
IB develops is a function of $x$, namely the larger $x$ is, the
larger $J_{c}/W$ becomes. Thus, we used Eq.~(\ref{RE}) to draw the
phase diagram $J_{c}/W$ vs. $x$ at various values of $V$. When
$V$=$0$ the occurrence of an IB due only to the $J$-term requires a
$J_{c}/W$$\approx$$0.25$ when $x$$\rightarrow$$0$, as seen in Fig.\
\ref{fig2}(a). When $x$$\rightarrow$$0$ and $J/W$$<$$0.25$ the
addition of a potential $V$ leads to the relation
$(J\smash{+}V)/W$$\approx$$0.25$ to establish the boundary of the
region where an IB develops. Our calculations also show that the
boundary between the IB and VB regions in the full $J$-$x$ plane
just moves down by an amount $\Delta(V)$ after the introduction of
the Coulomb attraction. This $\Delta(V)$ is independent of $x$
indicating that $J_c(x,V)= J_{c}(x,V\smash{=}0)-\Delta(V)$ as it can
be seen in Fig.\ \ref{fig2}(a).\cite{limit} This means that an IB
will be generated by a $J<J_c(x,V\smash{=}0)$ if a $V$ such that
$(J\smash{+}V)/W$$\approx$$J_{c}(x)/W|_{V\smash{=}0}$ is added.
Then, intuitively the effect of the addition of $V$ is to
renormalize $J$ to a larger value. This result is not surprising
because $J$ has a dual effect: \textit{(i)} it induces
ferromagnetism, but \textit{(ii)} it also tends to localize the
holes near the impurity so that they take advantage of the
antiferromagnetic coupling. This last property is similar to the
effect produced by the Coulomb attraction $V$. However, it would be
expected that as $x$ increases and more holes are added to the
system, the wave functions of the holes will start to overlap, and
as the holes become delocalized the effects of $V$ should become
less important.Thus, we would expect that the crossover boundaries
between the IB and VB regions indicated in  Fig.~\ref{fig2}(a)
should become closer to the $V=0$ curve as $x$ increases, instead of
remaining parallel as in the figure. Similar results have been
observed in MC simulations. \cite{yucel} We believe that the reason
for this unexpected behavior is related to the fact that here an
on-site central-cell potential is being considered. This behavior
can be corrected by considering a nearest-neighbor-range potential
\cite{yucel} or, within the DMFT framework, by considering a
phenomenological on-site potential that depends on $x$ such as
\begin{equation} \label{pot}
V(x)= V_{0} \exp{\{-(x/x_{0})^{2}\}},
\end{equation}
\noindent where $x_{0}$ can be roughly estimated
using Mott's criterion \cite{MOTT} as
\begin{equation}\label{x0}
x_0= \frac{0.25^3}{4}{\left(\frac{a_0}{a_B}\right)^{3}},
\end{equation}
with $a_0$ being the side of the cubic cell of the material and
$a_B$ the Bohr radius for the bound impurity. For a material such as
(Ga,Mn)As, which has an estimated $a_{B}\sim 8${\AA}, we obtain
$x_{0}$=0.0014. The resulting boundary between the IB and VB regions
is presented in Fig.\ \ref{fig2}(b) which indicates that for
realistic values of $J$ (0.2W) and $V_0$ (0.1W) for (Ga,Mn)As, the
crossover would occur for $x<0.5\%$.

After having remarked that some paradoxes of the results can be
solved by extending the size-range of the attraction or, similarly,
by reducing its strength with increasing $x$, here we will continue
the discussion of the qualitative aspects for the case of the
on-site central-cell potential. The main reason for it is to be able
to compare our conclusions with previous results in the literature
since an on-site potential is the only approach used in previous
numerical investigations.\cite{TAK03,HWANG05} There are still some
quantitative aspects that may need the $x$ dependent potential of
the previous paragraphs, and those will be clarified below.

Focusing on the on-site potential, it can be observed that even if
$J/W$$<$$J_{c}/W$, the IB regime can in general be reached either by
increasing $V$ at fixed $x$, or by decreasing $x$ at fixed $V$ (see
Fig.~\ref{fig2}(c)). While at $x$$\rightarrow$$0$ the carriers
trapped due to $V$ in the vicinity of each Mn core spin reside in an
impurity-like bound state, as $x$ increases the wave functions that
describe the bound state at the manganeses start overlapping (due to
the combined effects of $V$ and $J$) producing an IB that at a
critical $x_{c}$ merges with the VB. The renormalization condition
obtained in our calculations yields an IB-VB boundary in the diagram
$J/W$ vs. $V/W$, for a fixed $x$, as shown in Fig.\ \ref{fig2}(d).
This boundary deviates from linear only for very small values of
$J/W$ which is not a physically interesting region. According to the
results in Fig.\ \ref{fig2}(d) the area of the VB region is a
minimum for $x\rightarrow 0$ and increases with increasing $x$.

\subsection{Specific Results for (Ga,Mn)As and Other Compounds}

The literature does not provide a unique value of $V$ for the case
of (Ga,Mn)As. The main reason is that the value of $V$ necessary to
generate a bound state upon doping by one hole is a function of both
$J$ and the bandwidth $W$, as it can be observed from the results
presented in Table \ref{table}. Thus, in Ref.\ [\onlinecite{BAT00}]
a value of $V$ =2.3~eV is determined for $J$=0.9~eV with $W\approx
10~eV$ since a Luttinger-Kohn energy band is used, while in
Ref.~\onlinecite{TAK03}, $V$=0.6~eV is used with $J$=0.8~eV and
$W$=4~eV. In both cases, $V$ is determined by requesting that for a
single impurity doping a bound state at $E_{\rm b}$= 112~meV appears
as the combined result of the magnetic and Coulomb interactions. Our
calculations indicate that the parameters of Ref. \
[\onlinecite{BAT00}] provide an IB-VB crossover at $x_c\sim 0.5\%$
while we recovered the value $x_c\sim 3\%$ of Ref.\
[\onlinecite{TAK03}] using the parameters that they provided. The
discrepancy shows that the values assumed for $W$ and $J$ play an
important role in the determination of $V$ and $x_c$. The expression
given by Dietl,\cite{DIETL92} provides an estimation of the
non-magnetic impurity potential that may include more than Coulomb
interactions. It is evaluated using experimental data. For $x\approx
7\%$ \cite{HWANGEXP05,OKA99} with $W=3~eV$ and $J$=$1$ eV, the ratio
$|V/J|$=$0.55$ is obtained. The potential turns out to be repulsive
$V$=$-0.55$~eV. Notice that while the estimations of $V$ performed
for $x\rightarrow 0$ provides positive values, compatible with an
attractive potential, the estimations at finite doping do not. As
pointed out in the previous section, this indicates that it may be
necessary to use an $x$-dependent expression for the non-magnetic
interactions.

The phenomenological potential proposed in Eq.~(\ref{pot}) will
provide an IB-VB crossover at $x$$\sim$$0.1\%$ for all the
attractive values of $V$ provided above, as seen in
Fig.~\ref{fig2}(b).

We can make estimations of $x_c$ for (Ga,Mn)As and for other Mn
doped III-V materials as well. The value of $J$ is expected to be
inversely proportional to the volume of the cubic cell of the
material $a_0^{3}$, according to the chemical trends, and the energy
of the bound state for one Mn impurity has been
measured.\cite{DIETL02} From these data, we can estimate $V$ for
different values of $W$, with results given in Table \ref{table},
that also includes $a_0$ for each material and the estimated value
of $a_{\rm B}=\hbar/\sqrt{2m_kE_b}$ where
$m_k=m_e/(\gamma_1-(6\gamma_3+4\gamma_2)/5)$ with $m_e$ the electron
mass and $\gamma_i$ the Luttinger parameters.\cite{cardona} Then
$x_0$ can be obtained from Eq.~(\ref{x0}) and is also shown in the
Table. $x_c$ ($\tilde{x}_c$) indicates the estimated values of the
doping for which the IB-VB crossover occurs for an on-site
($x$-dependent) potential  ($V(x)$ given by Eq.~(\ref{pot})).
\vspace{-0.2in}
\begin{widetext}
\begin{center}
\begin{table}[h]
\caption{\label{Rep} DMFT calculated values of $V$ that
produce a bound state with energy $E_b$ for the values of $J$ and
bandwidth $W$ shown corresponding to the indicated DMSs. The
calculated doping density $x_{c}$ ($\tilde{x}_{c}$) at which the
IB/VB crossover occurs for an $x-$independent (dependent) potential
is listed. The IB label indicates that the material is in the IB regime
at all $x$$\in$$(0,1]$. Values of $a_0$, $a_B$, and $x_0$ (see text) for each
material are also shown.}
\begin{ruledtabular}\label{table}
  \begin{tabular}{cccccccccc}
    Material & J (eV) & $E_b$ (eV) & $W$(eV) & $V$(eV) & $x_c (\%)$&
$a_0$ (\AA) &$a_B$ (\AA) & $x_{0}$& $\tilde{x}_c$ (\%)\\ \hline
    (Ga,Mn)N
    &
    2.5
    &
    1.4
    &
    $\begin{matrix}
    10 \\  8 \\ 6 \\ 4
    \end{matrix}$
    &
    $\begin{matrix}
    2.7 \\  2.014 \\ 1.31 \\ 0.47
    \end{matrix}$
    &
    $\begin{matrix}
    \mathrm{IB} \\  \mathrm{IB} \\ \mathrm{IB} \\ \mathrm{IB}
    \end{matrix}$
    &
    4.42
    &
    1.6
    &
    0.082
    &
    $\begin{matrix}
     7.2 \\  9.3 \\ 21 \\ \mathrm{IB}
     \end{matrix}$\\ \hline
    (Ga,Mn)P
    &
    1.34
    &
    0.41
    &
    $\begin{matrix}
    10 \\  8 \\ 6 \\ 4
    \end{matrix}$
    &
    $\begin{matrix}
    2.4 \\  1.786 \\ 1.173 \\ 0.525
    \end{matrix}$
    &
    $\begin{matrix}
    5.2 \\  8.3 \\ 16.7 \\ 30
    \end{matrix}$
    &
    5.45
    &
    4.5
    &
    0.007
    &
    $\begin{matrix}
    0.422 \\  0.493 \\ 0.637 \\ 2.14\\
    \end{matrix}$\\ \hline
    (Ga,Mn)As
    &
    1.2
    &
    0.112
    &
    $\begin{matrix}
    10 \\  8 \\ 6 \\ 4
    \end{matrix}$
    &
    $\begin{matrix}
     1.883 \\  1.324 \\ 0.761 \\ 0.19
    \end{matrix}$
    &
    $\begin{matrix}
    0.52 \\  0.85 \\ 1.35 \\ 3.1
    \end{matrix}$
    &
    5.65
    &
    8
    &
    0.0014
    &
    $\begin{matrix}
    0.059 \\  0.068 \\ 0.09 \\ 0.37\\
    \end{matrix}$\\ \hline
    (Ga,Mn)Sb
    &
    0.96
    &
    0.016
    &
    $\begin{matrix}
    10 \\  8 \\ 6 \\ 4\\
    \end{matrix}$
    &
    $\begin{matrix}
    1.74 \\  1.232 \\ 0.698 \\ 0.175\\
    \end{matrix}$
    &
    $\begin{matrix}
    0.025 \\  0.045 \\ 0.064 \\ 0.13
    \end{matrix}$
    &
    6.10
    &
    39
    &
    0.00015
    &
    $\begin{matrix}
    0.00044 \\  0.00053 \\ 0.00065 \\ 0.0014
    \end{matrix}$
  \end{tabular}
\end{ruledtabular}
\end{table}
\end{center}
\end{widetext}
\vspace{-0.2in}

It is clear that for all relevant values of $x$, (Ga,Mn)As is in the
VB regime. The crossover, for realistic values of $W$, occurs at
$x$$\lesssim$$1\%$ for both on site and $x$-dependent potentials.
Thus, even including the Coulomb attraction, our results indicate
that the IB regime is not expected to play a relevant role in this
material. A similar picture emerges for (Ga,Mn)Sb. In this case the
IB-VB crossover is expected to occur for such small values of
impurity doping that for all practical purposes the Coulomb
attraction can be neglected.

On the other hand, the IB regime seems to dominate the physics of
(Ga,Mn)N. Considering $J$=$2.5$~eV, within our model we found that
even for the largest value of W considered (namely, $W$=10 eV) $J/W$
is strong enough to generate an IB region below some finite
$x_c(W)$, even if no Coulomb attraction is considered. However,
since the single hole bound energy for GaN is 1.4~eV, i.e. much
larger than the 0.113~eV value observed in GaAs, it is clear that
the Coulomb-attraction term has to be incorporated. In the table we
show the values of $V$ that together with $J$ will produce the bound
state for different values of the bandwidth $W$. Our calculations
show that with an on-site potential (Ga,Mn)N will be in the IB
regime for all relevant values of $x$ (we studied up to $x=80\%$).
This is still true when an $x$-dependent $V$ is considered since
even in the case for the largest bandwidth considered the crossover
is expected to occur at $x\approx 7.2\%$. Coulomb attraction should
therefore be included to study this material.

Our results for (Ga,Mn)P indicate that despite the deeper position of the
bound state in the gap, studies neglecting the Coulomb attraction could be
performed, particularly for $x\gtrsim 3\%$.

\begin{figure}[t]
{\scalebox{0.31}{\includegraphics[clip,angle=0]{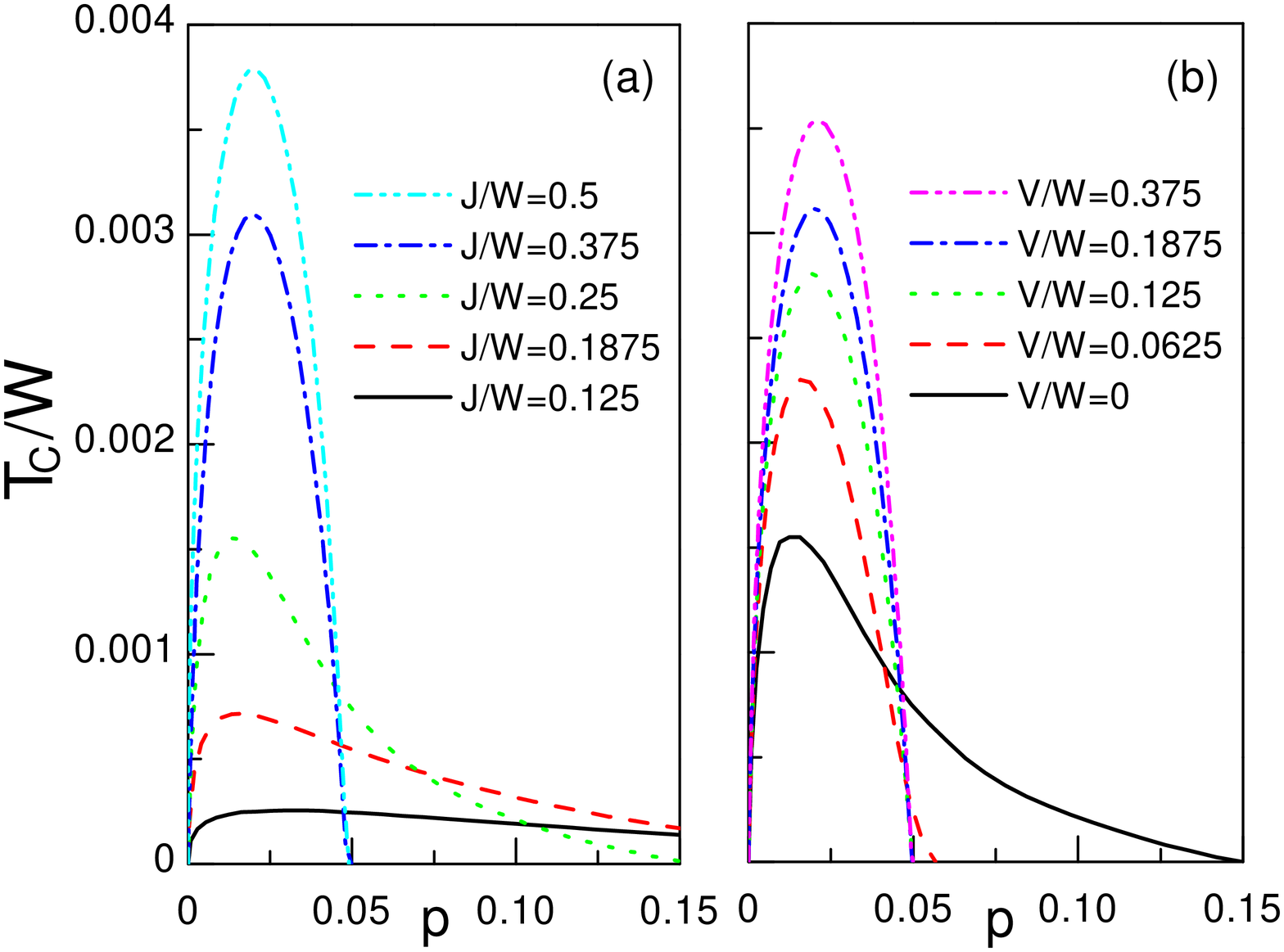}}}
\caption{(Color online) (a) $T_{\rm C}$ vs. $p$ at $V/W=0$ for
several values of $J/W$. (b) $T_{\rm C}$ vs. $p$ at $J/W$=$0.25$ for
various values of $V/W$. In both frames $x$=$0.05$.}\label{fig3}
\end{figure}
\begin{figure}[t]
{\scalebox{0.33}{\includegraphics[clip,angle=0]{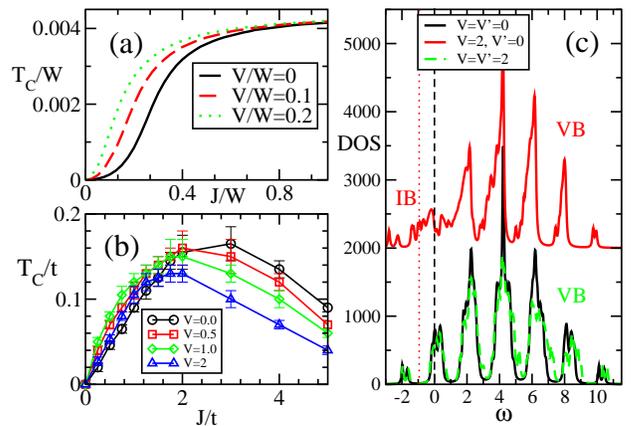}}}
\caption{(Color online) (a) $T_{\rm C}$ vs. $J/W$ at $p$=$0.015$ and
$x=0.05$ calculated with DMFT for different values of $V/W$. (b)
$T_{\rm C}$ vs. $J$ for different values of $V$ at $p_{h}$=$0.3$ and
$x=0.25$ obtained by MC. (c) The density of states (DOS) for $J/t=1$
and $V=0$ (black line); for an on-site Coulomb attraction $V=2$ (red
line; the curve has been shifted vertically for clarity); and for a
finite-range Coulomb attraction with on site intensity $V$ and next
nearest neighbors intensity $V'$=$V$=$2$ (dashed green line). The
vertical lines indicate the chemical potential. For clarity, the
curves for finite Coulomb attraction strength have been shifted
along $\omega$ so that the central peak in the DOS of all the curves
coincides.} \label{fig4}
\end{figure}

For completeness, and to compare with previous
calculations,\cite{TAK03} we present the $T_{\rm C}$ vs. $p$
dependence obtained from Eq.\ (\ref{TC}) at $x$=$0.05$, for
different values of $J$'s and no Coulomb attraction in Fig.\
\ref{fig3}(a). For $J/W$$\ll$$J_{c}/W$, $T_{\rm C}$ is low and
almost independent of $p$. When $J/W$$>$$J_{c}/W$, i.e. in the IB
regime, $T_{\rm C}$ vs. $p$ is semicircular with a maximum at
$p$=$x/2$, in agreement with previous results for one-orbital
models.\cite{nosotros} The behavior of $T_{\rm C}$ vs. $p$ at
different values of $V/W$ for $J/W=0.25$ is shown in Fig.\
\ref{fig3}(b). Comparing with the curves in part (a) of the figure
it is clear that $V$ increases the effective value of $J$. Our
results agree with Ref. \onlinecite{TAK03} and confirm that an
on-site square-well $V$ simply renormalizes $J$. The dependence of
$T_{\rm C}$ on $J$ for different values of $V$ is shown in Fig.\
\ref{fig4}(a). $V$ boosts $T_{\rm C}$ at small and intermediate
$J/W$, while at large $J/W$'s no change is observed because within
DMFT the $T_{\rm C}$ saturates as $J\rightarrow\infty$. However, as
it will be discussed in the following section, we believe that the
renormalization of $J$ for the physically relevant values of $x$,
such as the one used in our figures, is an artifact of the on-site
range of the Coulomb attraction and, thus, we do not expect it to
play a role in enhancing the $T_{\rm C}$ of real materials.

\subsection{Monte Carlo Simulations}

Hamiltonian (\ref{ham}) was also studied here using a real-space MC
technique with the Mn core spins treated classically. Details are
not provided since the technique has been widely discussed before in
the context of studies of manganites.\cite{DAG03} The simulations
were performed using cubic lattices with $4^{3}$ sites at
$x$=$0.25$. Finite-size effects have been monitored by running some
points on $5^{3}$ clusters. A random starting spin configuration has
been selected as the starting point for each temperature $T$. The
spins were allowed to evolve for a total of $10^{5}$ MC steps, with
the first $5$$\times$$10^{4}$ steps being discarded to thermalize
the starting configuration.

At $J/t$=$1$, and $p_{h}$=$p/x$=$0.3$, a value $V$=$1$ for the
on-site Coulomb attraction increases $T_{\rm C}$ by as much as 33\%,
as shown in Fig.\ \ref{fig4}(b). This agrees qualitatively with the
DMFT results. The figure shows clearly how $V$ effectively
"renormalizes" $J$. Since the curve $T_{\rm C}$ vs. $J$ for $V$=$0$
has a maximum at $J^{max}$ the effect of $V$ is to increase $T_{\rm
C}$ for values of $(J+V)$$\lesssim$$J^{max}$, while $T_{\rm C}$
decreases with $V$ for values of $(J+V)\ge J^{max}$. Although this
renormalization has been previously reported,\cite{TAK03} we do not
believe that it will play a role in the relevant range of doping for
most DMS. As we pointed out in subsection \ref{general}, the on-site
range of the Coulomb attraction induces unphysical behavior by
exaggerating hole localization for values of $x$ for which overlap
of the hole wave functions should occur. While finite range
attraction cannot be studied with DMFT, it can be done with MC
simulations but at the price of not being able to access the low
doping regime at which the IB-VB crossover would be expected to
occur for a material such as (Ga,Mn)As.

In Fig.\ \ref{fig4}(c), we present the DOS obtained with MC for
$J/t$=$1$ and $V$=$0$ for $x$=$25\%$ indicated by the black
continuous line. The peaks are due to the finite size of the system,
and each of them can be identified with the spikes that appear in
the DOS of a non-interacting system in the same lattice. Thus, at
this value of $J$, there is only a VB in the DOS, i.e., the magnetic
interaction is not strong enough to develop an impurity band. The
position of the chemical potential $\mu$ is indicated by the black
dashed line. Upon adding an on-site Coulomb attraction $V$=$2$, we
observe that an IB develops as indicated by the red line in the
figure, that has been shifted upwards along the vertical axis for
clarity. This IB is due to the localization of the holes induced by
the on-site potential. The chemical potential denoted by the dotted
red line indicates that only states in the IB are occupied. However,
when the range of the potential is increased to next-nearest
neighbors, as indicated by the green dashed line in the figure, it
can be seen that the IB dissapears although the intensity of the
potential has not changed. This occurs because, at this large
doping, the extended potential allows for a more uniform
distribution of the holes. As it can be seen in the figure, the DOS
for $V$=0 and for finite extended $V$ have an almost perfect
overlap. This shows that the use of on-site Coulomb attraction
potential can lead to missleading results and authors have to be
cautious when using this approximation.

\section{Conclusions}\label{Section5}

Our combined DMFT-MC study shows that the Coulomb attraction by
acceptors needs to be considered to obtain correctly the IB-VB
crossover as a function of impurity doping concentration $x$ in
models for DMS. However, for most materials we find that the
crossover occurs at very low levels of doping, outside the regime in
which high $T_C$ would be expected. We also find that a
doping-independent on-site square-well potential acts as a
renormalization of the coupling $J$ in an extended doping range up
to $x=80\%$. However, this apparent boost to the $J$-term at all Mn
dopings is unphysical, since the effect of $V$ should be
$x$-dependent beyond some critical value. Our MC simulations
demonstrate that this $x$-dependence is achieved naturally by
considering a longer range (next-nearest neighbors) square-well
attraction, which is beyond the capability of the single site DMFT
which can deal with on-site interactions only. Thus, a
phenomenological $x$-dependent Coulomb attraction was introduced.
With this modification, we have shown that for (Ga,Mn)As, the
Coulombic attraction $V$ influences the physics of the material only
at small Mn doping , i.e $x$$\lesssim$$0.5\%$. This result shows
that it is correct to  apply theories that consider the $J$-term
only for studying the  properties, including the Curie temperature,
of (Ga,Mn)As at the relevant values of Mn concentrations
$x$$\sim$$1\%$--$10\%$. On the other hand, we found that the Coulomb
attraction will play a relevant role, and should be included, in
studies of Mn-doped GaN.

Summarizing, here we have shown that the addition of an attractive
Coulomb potential is the necessary ingredient to explain the
transition from the IB to the VB regime as a function of Mn-impurity
doping concentration in materials for which the magnetic interaction
$J$ is not strong enough to bind a hole. However, we find that,
except for the case of (Ga,Mn)N, the crossover occurs at very low
doping in a regime in which high ferromagnetic critical temperatures
would not be expected and, thus, the effective value of $J$ will not
be affected. As a consequence, it is not necessary to include the
Coulomb attraction in the calculations. In addition, we show that an
on-site attractive potential does not capture the overlap of
localized hole wave-functions that should occur as a function of
doping and it provides unphysical results. Thus, to study materials
such as (Ga,Mn)N, in which the Coulomb attraction is relevant, a
nearest-neighbor finite range potential has to be used.

\section{acknowledgements}
We acknowledge helpful discussions with T. Dietl and J. Sinova. This
research was supported in part by the National Science Foundation
grants DMR-0443144 and DMR-0454504, and also in part by the Division
of Materials Sciences and Engineering, Office of Basic Energy
Sciences, U.S. Department of Energy, under contract
DE-AC05-00OR22725 with ORNL, managed by UT-Battelle.

\suppressfloats

\begin{references}
\bibitem{well} Ideally a potential with a finite range up to next nearest
neighbors should be used but its implementation is beyond the capabilities of
DMFT.
\bibitem{limit} The behavior displayed in Fig.~\ref{fig2}(a) was observed
up to $x\approx 80\%$, i.e., a percentage of impurity doping well
beyond the experimental posibilities. As $x$ approaches $100\%$ the
crossover is difficult to monitor because in this limit the system
would develop antiferromagnetic long range order and the valence
band is splitted in two sub-bands.
\bibitem{MOTT}  Mott criterion states that upon
doping a semiconductor a MIT occurs when the wave functions of the
holes localized around the impurities start to overlap. This occurs
when $n_{\mathrm{Mn}}^{1/3}a_{B}$$\sim$$0.25$. $n_{\mathrm{Mn}}$ is
the $\mathrm{Mn}$ density per unit volume and $a_{B}$ is the Bohr
radius of the hydrogenic impurity. Thus, we can obtain the relative
concentration $x_0$ of ${\mathrm{Mn}}$ for which the wave functions
would start to overlap in (Ga,Mn)As as
$x_{0}$=$n_{\mathrm{Mn}}/n_{\mathrm{Ga}}$, where the number of
$\mathrm{Ga}$ ions per unit volume $n_{\mathrm{Ga}}$=$4/a_{0}^3$.
Hence, Mott's criterion becomes
$(4x_{0}/a^{3}_{0})^{1/3}a_B$=$0.25$.
\end{references}
\end{document}